\pdfoutput=1

\documentclass[journal]{IEEEtran}

\usepackage{color}
\usepackage{graphicx}
\ifCLASSINFOpdf
\else
\fi
\usepackage{amsmath}
\usepackage{fancyhdr}
\begin{document}

\title{Optical Network Digital Twin\\ -- Practical Use Cases and Architecture}
%The following alternative title keeps the meaning staying in two lines
%\title{Optical Network Digital Twin - Commercialization Barriers, Use Cases, Value, and Challenges}

\author{Hideki~Nishizawa,~\IEEEmembership{Senior member,~IEEE, OPTICA} Toru~Mano, Kazuya~Anazawa, Tatsuya~Matsumura, Takeo~Sasai,~\IEEEmembership{Member,~IEEE, OPTICA}, Masatoshi~Namiki, Dmitrii~Briantcev, Renato~Ambrosone, Esther~Le~Rouzic, Stefan~Melin, Oscar~González-de-Dios, Juan~Pedro~Fernandez-Palacios, Xiaocheng Zhang, Keigo Akahoshi, Gert~Grammel, Andrea~D'Amico,~\IEEEmembership{Member,~IEEE,}
Giacomo~Borraccini,~\IEEEmembership{Member,~IEEE,} Marco~Ruffini,~\IEEEmembership{Senior member,~IEEE,}
Daniel~Kilper,~\IEEEmembership{Senior member,~IEEE,} Vittorio~Curri,~\IEEEmembership{Fellow,~IEEE,~OPTICA}
\thanks{H. Nishizawa, T. Mano, K. Anazawa, T. Matsumura, T. Sasai, and M. Namiki are with NTT Laboratories, Japan.
D. Briantcev, M. Ruffini, D. Kilper are with Trinity College Dublin, Ireland.
R. Ambrosone and V. Curri are with Politecnico di Torino, Italy.
E. Le Rouzic is with Orange Labs, France.
S. Melin is with Telia Company AB, Sweden.
O. Gonz´alez-de-Dios and J. P. Fernandez-Palacios are with Telefonica Innovación Digital, Spain.
X. Zhang is with NTT Docomo Business, Japan.
K. Akahoshi is with KDDI Research, Japan.
G. Grammel is with Juniper Networks, Germany.
A. D’Amico and G. Borraccini are with NEC Laboratories America, USA.
}
%
%\thanks{T. Mano is with NTT Network Innovation Laboratories, NTT Corporation, Yokosuka, Japan.}
%
%\thanks{K. Anazawa is with NTT Network Innovation Laboratories, NTT Corporation, Yokosuka, Japan.}
%
%\thanks{T. Matsumura is with NTT Network Innovation Laboratories, NTT Corporation, Yokosuka, Japan.}
%
%\thanks{T. Sasai is with NTT Network Innovation Laboratories, NTT Corporation, Yokosuka, Japan.}
%
%\thanks{M. Namiki is with NTT Network Innovation Center, NTT Corporation, Tokyo, Japan.}
%
%\thanks{D. Briantcev is with CONNECT Centre, Trinity College, Dublin, Ireland.}
%
%\thanks{R. Ambrosone is with Optical Communications Research Group, Politecnico di Torino, Torino, Italy.}
%
%\thanks{E. Le Rouzic is with Orange Labs, Lannion, France.}
%
%\thanks{S. Melin is with Telia Company AB, Solna, Sweden.}
%
%\thanks{O. Gonz´alez-de-Dios is with I+D/Global CTO Unit, Telefonica, Madrid, Spain.}
%
%\thanks{J. P. Fernandez-Palacios Gim´enez is with Innovación Digital, Telefonica, Madrid, Spain.}
%
%\thanks{X. Zhang is with Innovation Center, NTT DOCOMO Business, Tokyo, Japan.}
%
%\thanks{K. Akahoshi is with ***, KDDI Research, ***, Japan.}
%
%\thanks{G. Grammel is with ***, ***, München, Germany.}
%
%\thanks{A. D’Amico is with ***, NEC Laboratories America, Princeton, USA.}
%
%\thanks{G. Borraccini is with ***, NEC Laboratories America, Princeton, USA.}
%
%\thanks{M. Ruffini is with CONNECT Centre, Trinity College, Dublin, Ireland.}
%
%\thanks{D. Kilper is with CONNECT Centre, Trinity College, Dublin, Ireland.}
%
%\thanks{V. Curri is with Optical Communications Research Group, Politecnico di Torino, Torino, Italy.}
}

\maketitle

\begin{abstract}
With the widespread adoption of AI, machine-to-machine communications are rapidly increasing, reshaping the requirements for optical networks.
Recent advances in Gaussian noise modeling for digital coherent transmission have raised expectations for digital-twin-based operation.
However, unlike digital twins in wireless communication, which are already well established, significant barriers remain for commercialization in optical networks.
This paper discusses the evolving requirements of optical networks in the AI era and
proposes a practical Optical Network Digital Twin architecture enabling dynamic and Quality of Transmission aware operation beyond conventional management.
Representative use cases, including operator-driven optimization, user–operator collaboration, and multi-operator interconnection, are presented, along with the architectural framework and key challenges toward practical deployment.

\end{abstract}


% \fi
\IEEEpeerreviewmaketitle

\section{Introduction}
In recent years, the emergence of AI applications that consume massive, low-latency traffic has driven the deployment of large-capacity Data Center Interconnects (DCIs) in metropolitan areas. At the same time, the number of devices-to-device (M2M/IoT) connections is expanding rapidly.
While traffic has traditionally been dominated by human-to-human communication (e.g., mobile or FTTH), the growing demand for machine-to-machine and AI-enabled communication requires new optical network (ON) design and operational approaches.
For example, in AI model training, physical resources such as power, cooling, and floor space within a single facility often become limiting factors.
Leveraging multiple data centers can alleviate these constraints. \cite{nvidia2025turbocharge} reported training a 340-billion-parameter model distributed across two data centers approximately 1,000 km apart, each equipped with 1,536 GPUs.
Another example is broadcasting, where AI-driven innovation is transforming content creation and delivery.
With the advancement of Augmented and Mixed Reality, applications such as remote media production (RMP)~\cite{iowngf2024remote} are increasing, requiring flexible and high-capacity ONs.
Unlike conventional ONs based on long-term planning, the AI era demands rapid and flexible lightpath provisioning to accommodate on-demand and event-driven traffic.

Recent trends in optical transmission technologies further highlight this shift.
%
%The advent of digital coherent technology has driven substantial miniaturization and cost reduction of transceivers (TRxs), broadening their use beyond incumbent carriers to a wide range of network operators.
%
The introduction of digital technologies and the participation of multiple players in the industry have accelerated the standardization of transceiver (TRx) specifications and interfaces, thereby facilitating deployment in multi-vendor environments.
From a design-technology perspective, advances in additive white Gaussian-noise (AWGN) modeling have enabled near-real-time physical simulation of light signals, raising expectations for software-driven network design~\cite{curri2022gnpy}.
In the domain of propagation design, wireless communications—which historically adopted digital signal processing earlier than optical transport—has a well-established practice of simplifying E2E links between antennas and user devices and of applying digital-twin (DT) techniques to predict system-level traffic capacity and quality.
%
%For example, in~\cite{wireless_awgn}
The channel is modeled by extending an AWGN framework to incorporate fading, path loss, and interference, and uses these models to estimate communication capacity; field trials and urban-scale emulation in 5G/O-RAN environments have reported positive outcomes~\cite{owdt}, and industrial deployment have demonstrated measurable performance gains~\cite{ericsson_cht}.

\begin{figure}[ht]
  \centering
  \includegraphics[width=0.6\hsize]{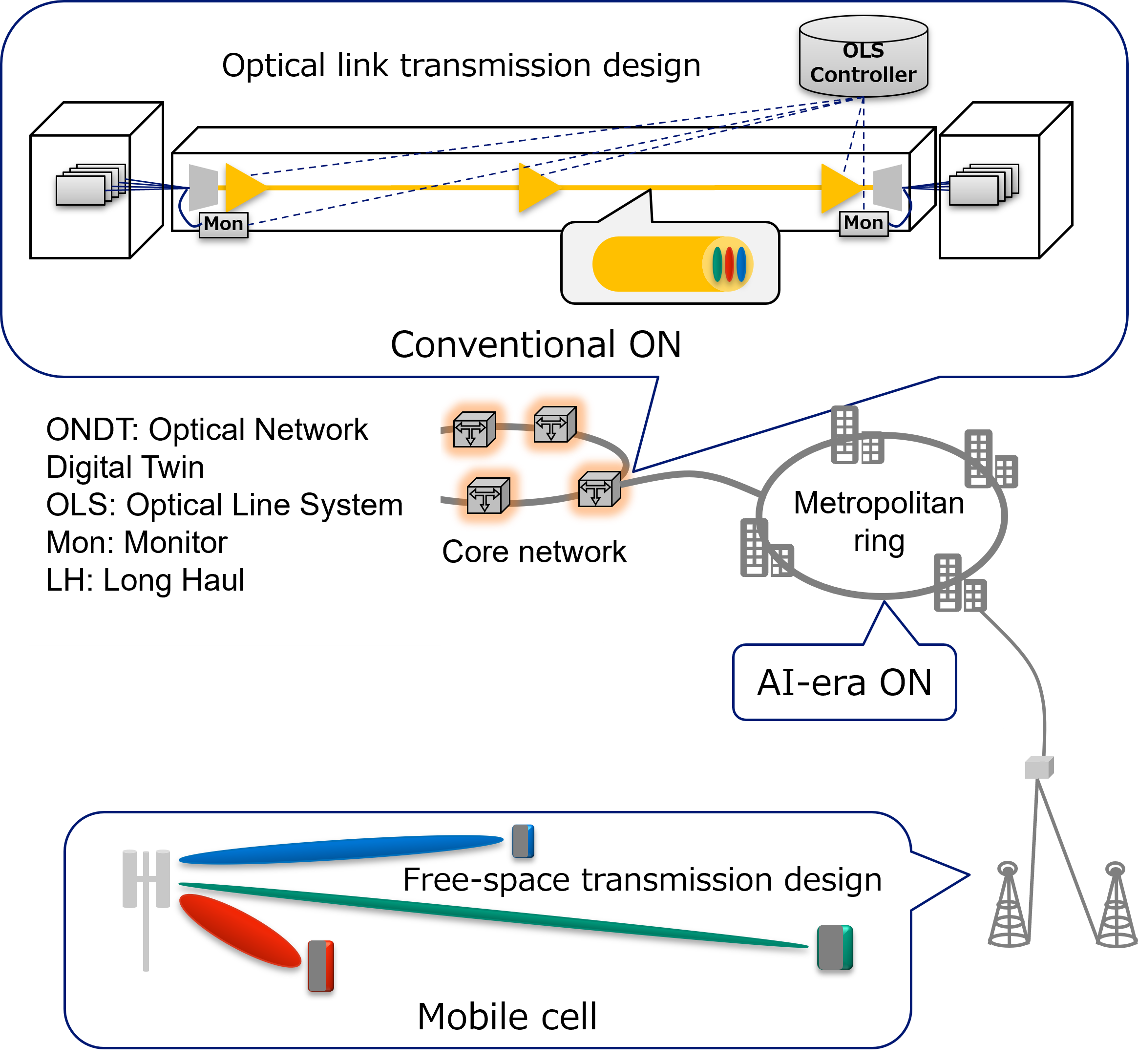}
  \caption{Conceptual comparison of optical networks and mobile systems}
  \label{optical_wireless}
\end{figure}

\begin{table}[t]
\caption{Comparison of operational characteristics}
\centering
%\footnotesize
\scriptsize
\setlength{\tabcolsep}{2.5pt}
\renewcommand{\arraystretch}{1.3}
\begin{tabular}{|p{1.3cm}|p{2.4cm}|p{2.4cm}|p{1.9cm}|}
\hline
\textbf{Item} & \textbf{Conventional ON} & \textbf{AI-era ON} & \textbf{Ref: Mobile cell} \\
\hline
Traffic variability & Statistical multiplexing (random traffic) & Specific patterns for AI applications & Statistical multiplexing \\
\hline
Number of users/link & $\sim$1,000,000 & 1$\sim$100 & 10$\sim$20 users/Cell \\
\hline
Link cost & Very large ($\sim$LH) & Large ($\sim$Metro) & Antenna only \\
\hline
Connection set-up time & A few months & $\sim$10 minutes & $\sim$Several seconds \\
\hline
Link usage time & 1 Month to 10 years & Months (AI training)
Hours (RMP) & $\sim$Minutes, hours \\
\hline
Link reliability & Strictly guaranteed as the lowest layer of carrier service infrastructure & Best effort or Guaranteed
(Redundancy in separate layers) & Best effort with multiple wireless services (4G/5G, WiFi) \\
\hline
Cost of failure & Compensation will be required based on an SLA with users. & Best efforts are not subject to compensation. & Not eligible for compensation as it is a best effort service. \\
\hline
Optimal operation method & Connection set-ups are required infrequently and must be extremely reliable, so manual operation is preferable. & Automatic operation is prevalent in DC, so the barriers to automation are low for optical connection set-up. & The number of link settings is enormous, so full automation is ideal. \\
\hline
\end{tabular}
\label{tab:comparison_ON}
\end{table}

In contrast, although TRx specifications and interfaces have become increasingly standardized, end-to-end (E2E) optimization methods and operational practices across optical devices—from TRx to the line system—remain not fully standardized.
Therefore, although automation technologies have advanced, realizing digital-twin-based operation that accommodates diverse user devices, as in wireless systems, remains a challenge in optical transmission.
To further discuss the requirements for ONs, TABLE~\ref{tab:comparison_ON} provides a comparative overview.
Traditionally, conventional ONs have formed the core segment of the infrastructure, supporting traffic from hundreds of thousands of users.
Since these networks often span long-haul distances, link costs are substantial, setup may require several months, and the operational lifetime typically extends to around ten years.
To honor the numerous service level agreements (SLAs) concluded with their customers, carriers must construct ONs with extremely high reliability.
%often targeting up to 99.9999\% availability.
%
%For this reason, manual operation, although benefiting from extensive remote control automation, has become the usual practice, constituting a significant barrier to more wide scale deployment.
%
For this reason, automation in ON operation remains less advanced than in wireless systems and data center environments.
Although a direct comparison is challenging, it is informative to consider mobile cells, where research on DT techniques has already been actively explored, as a reference case.
Mobile cells are located in the user access portion of the network, typically supporting on the order of tens of simultaneous connections per cell.
Link setup is completed within a few seconds, and link utilization generally lasts from minutes to hours.
As a best-effort service, there is no compensation to users in the event of link failure. 
Due to the vast number of link establishments combined with relatively modest reliability requirements, automation has proven readily applicable in this context.
Although the AWGN model is applicable to digital coherent communications—similarly to wireless systems—the application of advanced automation technologies to commercial ONs remains challenging.

Related research is summarized as follows.
\cite{vilalta2023ndt} proposed and evaluated an architecture applying DT concepts to ONs using a cloud-native software-defined network (SDN) controller.
Since ONs exhibit complex physical-layer characteristics that are partially inaccessible, accurate modeling and prediction remain difficult. The authors therefore highlighted the importance of DT application to optical domains and outlined potential use cases.
%not fully accessible to the planner, it is difficult to model and predict their behavior accurately.
%
%Accordingly, the authors emphasized the significance of applying network DT techniques to optical domains and presented several potential applications.
%
In~\cite{borraccini2023experimental}, a multi-vendor SDN architecture integrating a physical-layer DT within a intent-based control framework was demonstrated. 
The experimental proof of concept demonstrated E2E ON, including ROADMs, in-line amplifiers, and whitebox transponders with pluggable multi-rate TRxs.
{
In this context, an Optical Network Digital Twin (ONDT) is defined as a near real-time digital representation that integrates physical-layer modeling and monitoring information to estimate and visualize E2E Quality of Transmission (QoT) for each light path, along with a feedback mechanism to reflect these results in network operation.
By leveraging AWGN-based models and TRx-based measurements, ONDT enables computationally efficient QoT estimation, supporting software-driven design, provisioning, and operation of ONs.
Unlike conventional offline planning tools, ONDT continuously reflects the network state and enables closed-loop, QoT-aware control, where control decisions are fed back to and applied in the physical network.

While individual techniques have been studied in isolation, this paper presents a unified framework that captures their evolution and integration into a practical ONDT architecture. The contributions of this paper are summarized as follows:

\subsubsection{Evolution of optical network operation}
We present a three-step evolution of ON operation toward the AI era and discuss how ONDT-based operation enables new operational capabilities at each stage:

\begin{itemize}
\item \textit{Operator-driven optimization:} It enables cost reduction and flexible service-level configuration through freedom of device selection and minimization of unnecessary margins, allowing operators to adapt transmission performance to application requirements, supported by QoT visualization and real-time performance awareness.

\item \textit{User--operator collaboration:} It supports on-demand lightpath design and optimization between arbitrary locations, including protocols and required operation time based on prior experimental and implementation results.

\item \textit{Operator--operator collaboration:} It enables multi-operator interconnection across domains through coordinated operation between operator controllers, while highlighting the importance of efficient utilization of optical spectrum across operators.
\end{itemize}

\subsubsection{ONDT architecture with managed/unmanaged OLS models}
We extend the ONDT architecture by introducing and structuring two types of optical line system (OLS) models---managed and unmanaged---which have not been fully addressed in prior work and enable efficient utilization of existing  optical infrastructures.
%
%While various Optical Network Digital Twin (ONDT) architectures and validations have been reported, their operational value and practical use in commercial networks remain unclear.
%
%The paper focuses on the evolving requirements for ONs in the AI era and clarifies two points from an operator’s perspective.
%
%\begin{itemize}
%    \item By utilizing an end-to-end (E2E) physical model based on the AWGN, applying visualization, and using the bit error rate (BER) as a QoT parameter estimator for feedback, a simple yet accurate implementation of an ONDT can be achieved.
%    \item By leveraging the above approach, E2E quality management and advanced network service provisioning per optical path becomes feasible, enabling a stepwise adaptation to the evolving performance and flexibility requirements of ONs in the AI era.
%\end{itemize}

Following the discussion of commercialization barriers presented in TABLE~\ref{tab:comparison_ON}, Sec. II presents the practical use cases, Sec. III discusses the architecture for ONDT, and Sec IV-V outline future challenges and conclusions.
}
%Following the discussion of commercialization barriers in this section, Sec.~\ref{sec:ONDT_archi} discusses the architecture of the proposed ONDT and the values it delivers. 
%
%Sec.~\ref{sec:early_use_cases} presents early use cases and the evolutionary steps.
%
%Sec.~\ref{sec:QoT_visualization} introduces experimental results on QoT visualization and its accuracy, and Sec.~\ref{sec:future_challenges} concludes with future challenges.
{

\begin{figure*}[ht!]
  \centering
  \includegraphics[width=0.9\hsize]{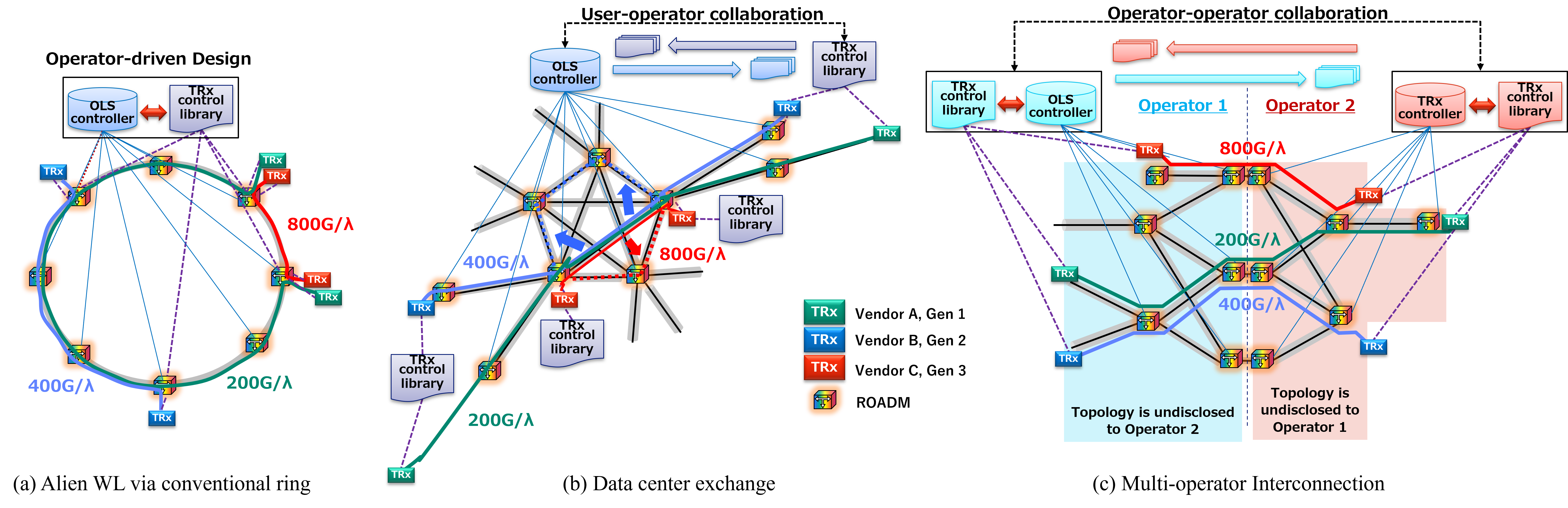}
  \caption{Practical use cases for Optical network digital twins}
  \label{practical_usecase}
\end{figure*}

\section{PRACTICAL USE CASES}
\label{sec:practical_use_cases}
Fig.~\ref{practical_usecase} illustrates representative practical use cases of ONDT in metro ONs.
The figure highlights three key scenarios: (a) multi-generation interoperability in alien wavelength deployments, (b) data center exchange with topological flexibility and on-demand provisioning, and (c) multi-operator interconnection across domains.
These use cases reflect the evolving operational requirements of ONs, ranging from operator-driven optimization to user–operator and operator–operator collaboration.
In the following subsections, we discuss each scenario and its associated value proposition in detail.
\subsection{Multi-Generation Interoperability in Alien Wavelength Scenarios}
\label{Alien_WL}
As illustrated in Fig.~\ref{practical_usecase}(a), in alien wavelength deployments over point-to-point (P2P) or ring networks, TRxs from different vendors and generations must interoperate over diverse optical infrastructures.
Such disaggregated TRx are exemplified by solutions such as TIP MANTRA IPoDWDM and Phoenix, project which promote vendor-agnostic optical interfaces.
While multi-generation interoperability within a single-vendor system has already been demonstrated in commercial deployments and can be highly optimized using vendor-provided design tools and models, such approaches inherently rely on vendor-specific assumptions and limited transparency.
As a result, they do not readily extend to multi-vendor environments or operator-driven optimization, where consistent and transparent QoT evaluation across heterogeneous systems is required.
In addition, existing optical line systems (OLS), such as P2P and ring networks, often contain unused wavelength channels~\cite{Kaeval2025} due to long infrastructure lifecycles. While OLS deployments are typically designed for operation over periods of up to ten years, TRx technologies evolve much faster, with new generations introduced every few years. As a result, higher-rate TRxs (e.g., 400G/$\lambda$ and 800G/$\lambda$) may become available before all wavelength channels are fully utilized.

ONDT enables an operator-driven approach, where interoperability across multi-vendor and multi-generation TRxs is determined based on real-time QoT estimation rather than vendor-specific assumptions.
This allows operators to exploit unused wavelength channels in existing OLS and deploy higher-rate TRxs even in systems originally designed for lower per-wavelength capacities (e.g., 200G/$\lambda$).
For example, replacing four 200G TRxs with a single 800G TRx can reduce both capital and operational costs. In general, the cost of an 800G TRx is approximately 2.5-3.5 times that of a 200G TRx, rather than scaling linearly with capacity.
As a result, consolidating four 200G TRxs into one 800G TRx can reduce the total TRx cost by approximately 20–40\%, while also significantly decreasing power consumption and footprint.
In addition to minimizing unnecessary margins, ONDT enables QoT-aware service-level adaptation by adjusting transmission parameters according to application requirements.
For example, services requiring extremely high reliability (e.g., RMP) can be provisioned with sufficient QoT margins, whereas bandwidth-oriented services (e.g., large-scale data backup) can operate with reduced margins to maximize spectral efficiency.
Such dynamic trade-offs between reliability and efficiency are determined based on real-time QoT estimation, allowing operators to flexibly tailor service levels. This represents a shift from static, worst-case design to operator-driven, QoT-aware service control.
Furthermore, ONDT enables interoperability across vendors (bookend, or partially disaggregation), allowing operators to adopt not only vendor-native TRxs but also third-party TRxs.
This reduces vendor lock-in and increases procurement flexibility, potentially leading to additional cost reductions.
Such optimization becomes feasible with ONDT, which enables accurate QoT estimation and minimizes unnecessary design margins, allowing higher-rate and multi-vendor TRxs to be deployed over existing infrastructure.
This demonstrates that ONDT can unlock latent capacity and cost efficiency in existing optical infrastructures.
\subsection{Data Center Exchange with Topological Flexibility and On-Demand Provisioning}
\label{sec:DCX}
As shown in Fig.~\ref{practical_usecase}(b), in data center interconnection scenarios, users and applications distributed across multiple data centers dynamically request connectivity depending on workload placement and traffic demand.
Such on-demand light path provisioning has been demonstrated in prior work. In particular, a cooperative protocol for light path design between users and operators has been proposed, enabling dynamic provisioning over alien access links through QoT-aware configuration and parameter exchange~\cite{Nishizawa2023dynamic}. 
Field trial results have demonstrated that light paths can be automatically established within approximately several minutes (e.g., within about 6 minutes), based on AWGN-model-based QoT estimation, confirming the practical feasibility of on-demand provisioning.
In addition, when the carrier remotely deploys and installs a control container on the user transponder to execute the required sequence, an additional 90 seconds is required~\cite{nishizawa2024fast}.

In practical operation, the time required to establish and optimize an E2E light path depends on network conditions. When no significant impairments are present, QoT can be estimated from BER measurements at the TRx, and the provisioning and optimization process can typically be completed within approximately 10 minutes.
In cases where performance degradation is observed and further investigation is required, additional diagnostics such as power profile verification using Digital Longitudinal Monitoring (DLM) may be performed. For example, DLM-based estimation over a 5-span (198.3 km) link has been demonstrated to be completed in approximately 7.5 minutes~\cite{Nishizawa2024semi-automatic}, resulting in additional time overhead depending on link conditions. 

However, previous studies primarily focus on establishing individual connections and do not fully exploit global network optimization.
An ONDT-enabled DCX architecture extends this capability by enabling rapid QoT estimation for each E2E light path between arbitrary nodes using an AWGN model. This allows efficient evaluation of multiple candidate routes and transmission configurations across the network.
When a new connection request is triggered, the ONDT-enabled DCX architecture dynamically selects feasible paths and configures TRxs based on current network conditions. In addition, the system continuously re-evaluates existing connections and adapts routing and transmission parameters as traffic patterns evolve.
This enables not only on-demand provisioning but also continuous optimization of network resources, where routing, capacity, and QoT margins are jointly optimized.
As a result, the ONDT-enabled DCX architecture supports efficient and flexible data center exchange across distributed environments, moving from per-connection provisioning toward a globally optimized and operator-driven network operation.
\subsection{Multi-Operator Interconnection Across Domains}
\label{sec:Multi-Operator_Interconnection}
As depicted in Fig.~\ref{practical_usecase}(c), in multi-operator environments, E2E optical connectivity across multiple domains requires coordination among independent operator control systems while preserving domain autonomy and confidentiality.
An inter-operator coordination framework supported by ONDT-based modeling enables consistent E2E QoT estimation across domain boundaries without requiring full disclosure of internal network details. By leveraging abstracted physical-layer information and model-based QoT evaluation, feasible transmission configurations can be determined across heterogeneous domains.
This capability enables coordinated operation between operator controllers, allowing operators to establish E2E light paths spanning multiple domains while maintaining control over their respective infrastructures. Compared with conventional approaches relying on strict demarcation and conservative margins, such coordination improves the utilization of optical spectrum and enhances flexibility in service provisioning.
Quantitative studies on multi-operator optical spectrum sharing have demonstrated significant benefits, including up to 3.5× CAPEX reduction, up to 6.5× improvement in power efficiency, and noticeable reductions in E2E service downtime~\cite{Kaeval2025}.
These results indicate that ONDT-based operation can significantly improve the efficiency and scalability of multi-operator ONs.
}
\section{Optical network digital twin architecture}
\label{sec:ONDT_archi}
This section presents the ONDT architecture that enables practical use cases and discusses its application depending on whether the QoT of the Optical Line System (OLS) is managed or unmanaged.
%
%This section introduces the proposed ONDT in terms of its underlying modeling and a QoT visualization example.
%
%We also discuss the differences in the approach to applying ONDT, depending on whether the QoT of the Optical Line System (OLS) is under management or not.

\subsection{Digital Twin Model}
\label{subsec:dt_modeling}

\begin{figure}[ht!]
  \centering
  \includegraphics[width=1.0\hsize]{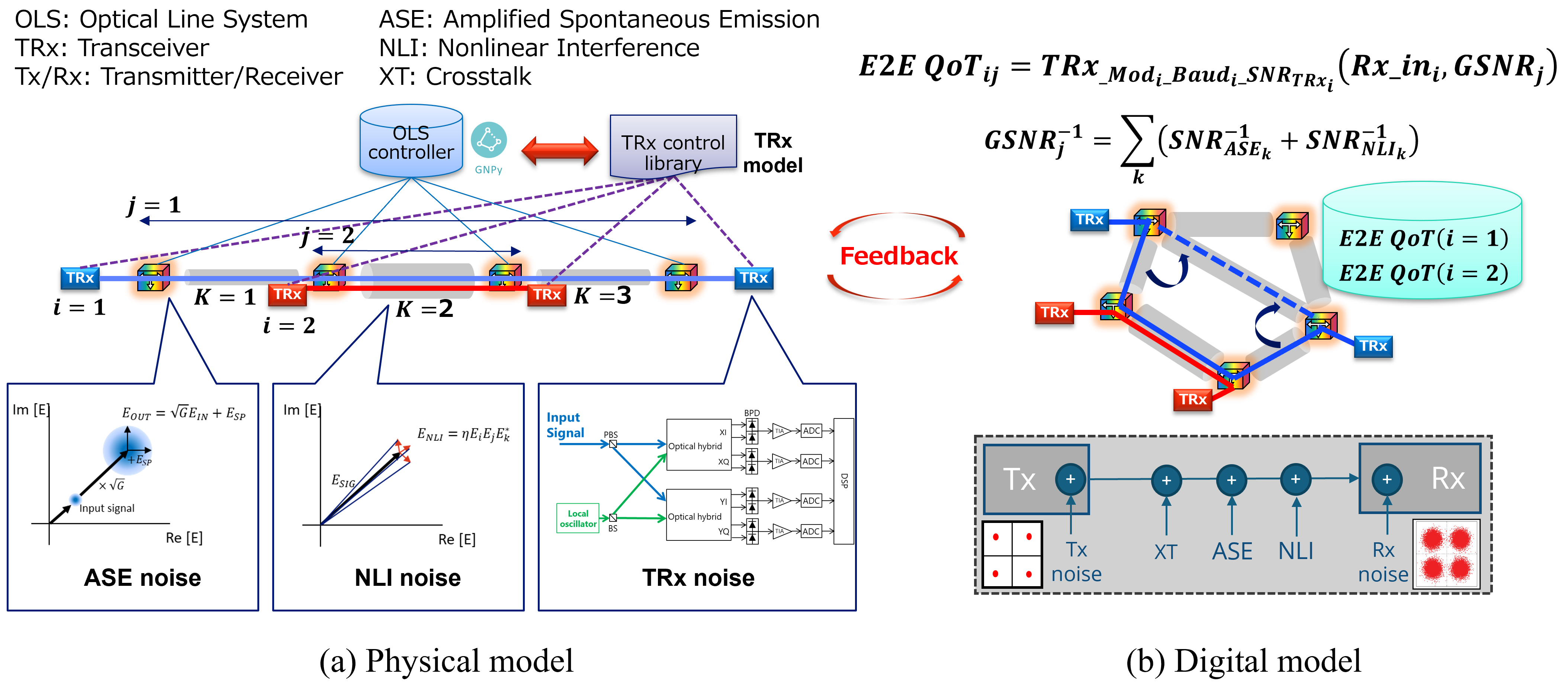}
  \caption{Optical network digital twin architecture}
  \label{ondt_archi}
\end{figure}

Fig.~\ref{ondt_archi} (a) and (b) shows the physical and digital models of the proposed ONDT architecture, respectively. While similar offline planning tools have been used for some time, the approach here is novel as the ONDT is maintained in real time and used for online network control and management.
Based on the AWGN assumption, the OLS QoT estimation can be obtained evaluating the generalized signal-to-noise ratio (GSNR) that serves as characterizing metric of the OLS transmission impairment. 
In this framework, accurate algorithms for OLS QoT evaluations have been already implemented in GNPy~\cite{curri2022gnpy}, accounting for the main contributions to the GSNR: the signal power profile, the amplified spontaneous emission (ASE) noise and the nonlinear interference (NLI) noise, evaluated along the OLS components.
The signal power profile is computed element by element along the light path, taking into account fiber attenuation, erbium-doped fiber amplifier (EDFA) gain, and, when relevant, gain or power saturation effects.
For the TRx, it has been verified that GN-model-based TRx models, proposed and discussed within the IOWN Global Forum, can accurately reproduce transceiver behavior~\cite{mano2023modeling}.
For a suitable model, the TRx characteristics, such as BER vs. OSNR and forward error correction (FEC) limits as well as QoT information (e.g., real-time pre-FEC BER), need to be measured and collected~\cite{nishizawa2024fast}. 
%The TRx controller accesses the transceiver to obtain its characteristics (such as BER vs. OSNR and FEC limits) as well as QoT information (e.g., real-time pre-FEC BER)~\cite{nishizawa2024fast}.
%
Optical transmission in metro-areas is primarily impaired by TRx noise, ASE noise, and NLI noise.
It has already been demonstrated through field-fiber experiments that these impairments can be accurately estimated using the AWGN model~\cite{mano2023modeling, nishizawa2024fast}.
The E2E QoT of each TRx over a given link is determined by the received optical power and the GSNR, while TRx parameters such as modulation format, baud rate, and intrinsic noise are treated as fixed characteristics.
Assuming a GN-model-based approximation, the GSNR is obtained by accumulating noise contributions along the link, including ASE and NLI, as illustrated in Fig.~\ref{ondt_archi} (b).

%The E2E QoT of $TRx_i$ over $link_j$ is modeled as a function of the received optical power at $Rx_i$ and the $GSNR_j$, with the modulation format $Mod_i$, baud rate $Baud_i$, and TRx noise $SNR_{TRx_i}$ considered constant parameters.
%
%Here, assuming an incoherent GN model, the $GSNR_j^{-1}$ is expressed as the sum of the $SNR_{ASE_k}^{-1}$ and $SNR_{NLI_k}^{-1}$ contributions accumulated along $link_j$~\cite{nishizawa2024fast}. 

In the digital domain described in Fig.~\ref{ondt_archi}(b), the E2E QoT for each link is calculated based on the formulation shown in the figure and stored in a database together with the network topology information, and QoT management is performed on a per light path basis rather than per WDM link. 
When a user requests a new path setup, the potential impact on the existing users’ E2E QoT is computed within the digital space.
Here, the operators themselves can design the margin for each light path according to the user’s requirements.
The proposed ONDT architecture integrates QoT monitoring and historical data within a unified digital representation of the network, maintaining consistency between the physical and digital domains.
Because this digital representation continuously reflects the network state, the ONDT enables the evaluation of network conditions and alternative configurations in advance, supporting rapid and informed operational decisions in response to potential degradations or failures.

%This architecture also contributes to maintenance.
%
%By monitoring the QoT of many E2E light paths and storing their historical data, it becomes possible to localize fault points and enable proactive fault management when quality degradation occurs on a particular link.
%
%Furthermore, to prepare for possible failures or malfunctions, multiple backup routes are precomputed in advance before any faults occur in the digital domain.

\textbf{QoT visualization}:

A key function of ONDT is to transform physical-layer performance metrics into intuitive and actionable information, enabling operators to understand, control, and optimize E2E light path quality in near real time, with control decisions continuously reflected back to the physical network, forming a closed-loop operation.
Here, such visualization is critical from the perspectives of compatibility with existing operations, transparency, and explainability, particularly in complex multi-vendor environments.
%
%Taking autonomous driving as an example, the automation level progresses step by step: from Level 1, where humans are involved, toward Level 5, which is full automation.
%
%Moreover, in the event of an accident, it is essential to be able to trace and explain the cause.
%
Since data center infrastructures are vital social foundations, it is necessary to advance automation while ensuring transparency and gaining public trust.

\begin{figure}[ht!]
  \centering
  \includegraphics[width=1.0\hsize]{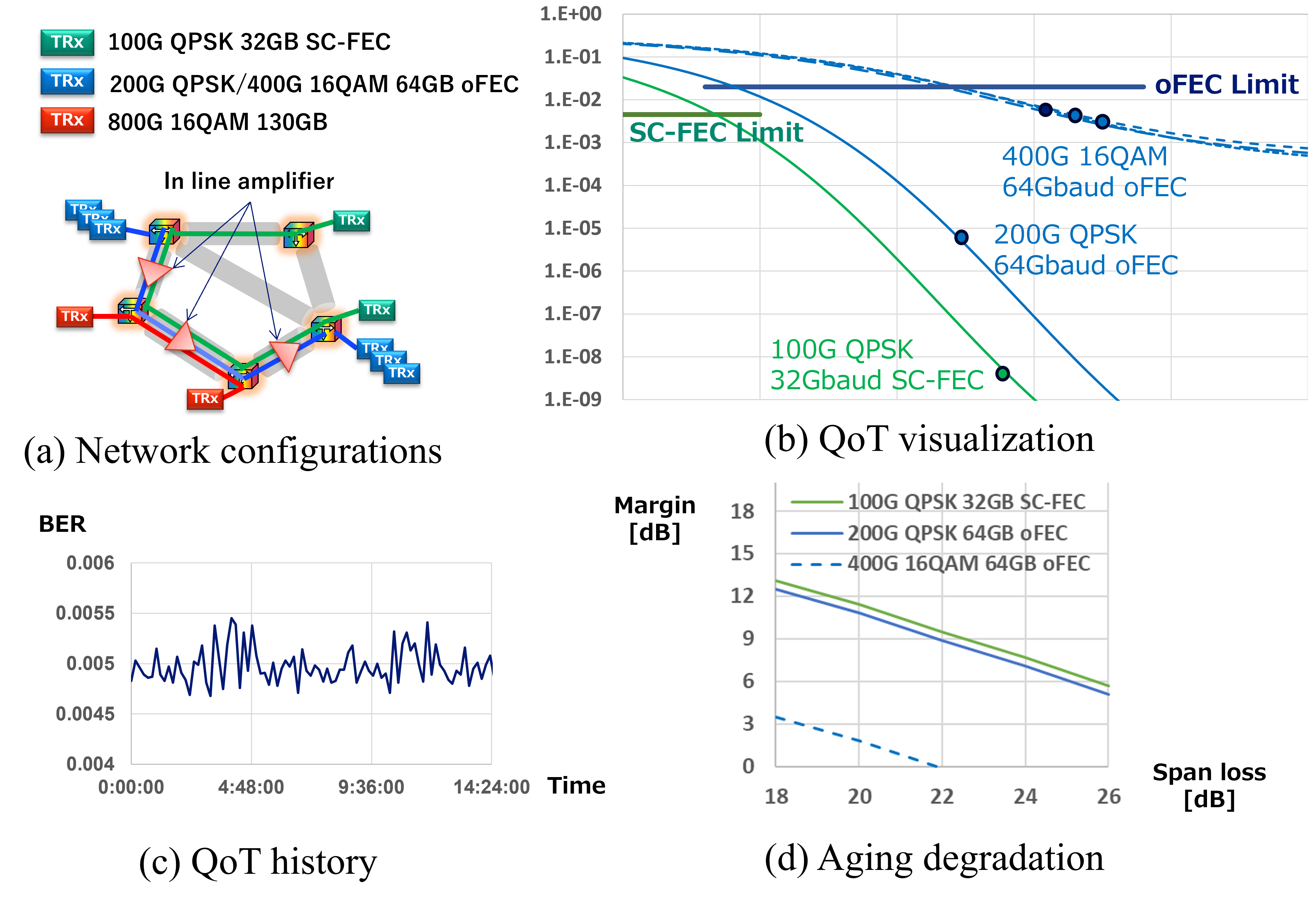}
  \caption{QoT visualization}
  \label{qot_visualizatin}
\end{figure}
Fig.~\ref{qot_visualizatin} presents the architecture shown in Fig.~\ref{ondt_archi} from an operator’s perspective, providing a clearer visualization of its practical configuration.
The network topology is depicted in Fig.~\ref{qot_visualizatin}(a), where three generations of TRxs are utilized.
%: 100 Gbps QPSK 32 Gbaud with SC-FEC, 200 Gbps QPSK 32 Gbaud/400 Gbps 16QAM 64 Gbaud with oFEC, and 800 Gbps 16QAM 130 Gbaud with proprietary FEC.
%
These TRxs interconnect arbitrary N-to-N sites with different source and destination points through ROADMs and inline amplifiers.
In (b), the E2E QoT of each TRx shown in (a) and the FEC limits (standardized) are mapped onto the BER-vs-OSNR curves.
%
%Under the assumption that the primary impairments in metro-area digital coherent transmission are noise contributions from the TRx, ASE, and NLI, the intersection between the measured BER value and the back-to-back characteristic corresponds to the GSNR.
%
The characteristics of a TRx can be represented by its BER-vs-OSNR curve and FEC limit, where the margin can be visually identified from the intersection of these two curves and the corresponding GSNR value.
Such visualization enables operators to centrally manage the quality of multiple TRxs, regardless of differences in their characteristics, generations, or the E2E light paths they traverse.
As reported in~\cite{nishizawa2024fast}, the measurement accuracy of this method was experimentally verified.
%using field-deployed fibers with several ROADMs installed in urban areas and two types of multi-vendor TRxs with different types of modulators.
%
%The modeling error was found to be within the same order as the inherent line-system error of 0.2 dB.
%
The modeling error was found to be on the same order as the typical measurement uncertainty of practical line systems (approximately 0.2 dB)
Fig.~\ref{qot_visualizatin}(c) presents an example of QoT history. We measured BER over night with three polarization controllers.
Fig.~\ref{qot_visualizatin}(d) illustrates the degradation of the Q-factor when equal losses are inserted into each span of the line system, representing quality deterioration caused by route changes due to fiber cuts or by water ingress.
Leveraging the digital model enables straightforward simulation of such degradation over time. Moreover, adopting the open-source GNPy as a common reference and using BER (defined in standard interfaces such as Centum gigabit Form factor Pluggable Multi-Source Agreement (CFP MSA) management interface specification and Optical Internetworking Forum (OIF) Common Management Interface Specification (CMIS)) as a QoT parameter allows operators to ensure network reliability and provide services even in multi-vendor environments.

\subsection{OLS types (Managed/unmanaged)}
\label{managed_unmanaged_ols}
\begin{figure}[ht!]
  \centering
  \includegraphics[width=1.0\hsize]{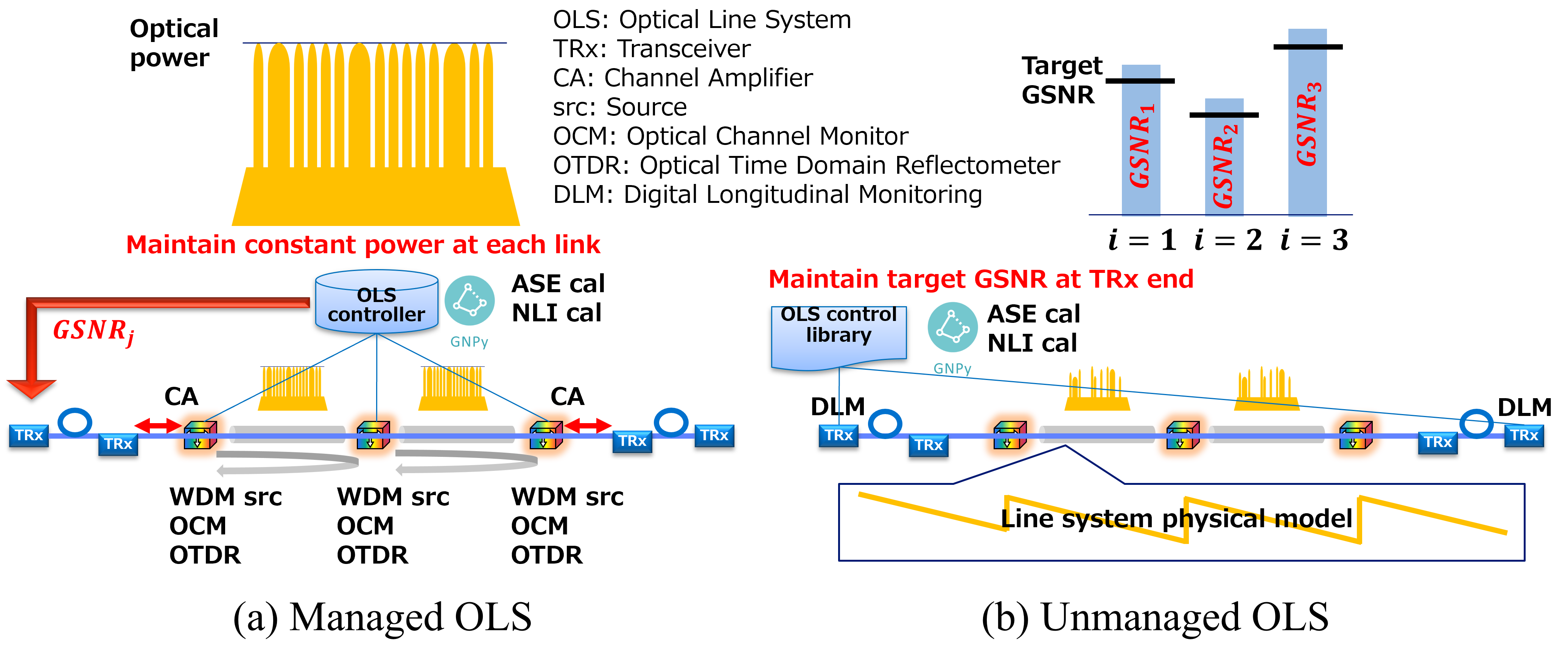}
  \caption{Changes in OLS types and operational methods}
  \label{ols_type}
\end{figure}

Fig.~\ref{ols_type} represents two different OLS types where the TRx is remotely connected to the OLS via an access link, and the optical input power to the OLS differs depending on the loss of the access link and variations in the optical output power of the TRx.
These OLS types broadly correspond to partially and fully disaggregated architectures, respectively.
(a) shows an example of the managed OLS,  
Each ROADM node has monitor functions such as Optical Channel Monitor (OCM) and Optical Time Domain Reflectometer (OTDR).
The signal power and GSNR of each wavelength path are monitored and constantly optimized in a link-by-link manner within the OLS. 
Furthermore, by utilizing a WDM source to inject pseudo signals into unoccupied channels, high-accuracy power management can be achieved. 
Edge ROADMs have Channel Amplifiers (CAs) connected to the remote TRxs to optimize power transmission between the ROADMs and the remote TRxs. 
In this case, $GSNR_j$ is calculated by the OLS controller where the power of each wavelength is controlled to remain constant for every link.

Fig.~\ref{ols_type}(b) shows an example of the unmanaged OLS.
The power and GSNR of each wavelength path are not monitored or controlled link-by-link within the OLS. 
Monitoring functions such as Digital Longitudinal Monitoring (DLM) and the OLS control library included~\cite{Nishizawa2024semi-automatic} in GNPy are implemented near the TRxs to calculate $GSNR_j$. 
DLM is a viable tool for network monitoring, as it enables visualization of E2E, distance-resolved optical power solely from coherent receivers.
In addition, it facilitates the identification of performance bottlenecks in multi-domain networks and enables fault detection along optical transmission paths.
Its feasibility has been verified in live production networks under multi-domain environments~\cite{sasai2025production}.
Enabling the TRx to estimate line-system parameters in this manner allows operators to ensure consistent transmission quality.
In particular, when E2E light paths are established between arbitrary nodes using TRxs of different generations and the QoT is managed on a per light path basis, the configuration shown in (b) proves more effective, offering improved service flexibility and quality assurance with reduced operational cost.
In operating such a network, it is no longer critical to maintain constant power per channel.
Rather, as illustrated in the upper part of (b), network monitoring focuses on ensuring that the E2E light path meets its required target GSNR.
Since the configuration in (a) and similar variants is currently the most common, service deployment will initially start with (a).
However, with advancements in monitoring technologies and ONDT-enabled automation, ONs are expected to evolve toward the approach represented in (b), particularly in the AI era.

\section{Future challenges}
\label{sec:future_challenges}
The proposed physical model includes TRx, ASE, and NLI noise; however, the margin can be further reduced by incorporating other impairments. For example, in the case of PDL induced by the WSS in a ROADM, no ideal model has yet been established, and the system is therefore operated with excessive margin.
In addition, linear crosstalk induced by the WSS is also an attractive research topic.
While this study focuses on metro networks, additional effects must be considered to extend the approach to long-haul transmission. TRx-based monitoring techniques such as DLM are expected to expand measurable parameters and enhance accuracy.
For unmanaged OLSs, future work should also address scenarios with numerous wavelengths, gain variations among channels, and power fluctuations caused by sudden channel shutdowns.
In multi-operator environments, defining interfaces and demarcation points, as well as ensuring security and management of shared line-system information, will be essential.

In addition, appropriate abstraction of QoT-related parameters and coordination mechanisms between operator controllers will be required to enable scalable multi-domain operation.
Finally, research on dynamic optical-frequency allocation and linkage with upper-layer orchestrators will be important to realize fully automated ONDT-based operation.
Such mechanisms are also closely related to efficient utilization of optical spectrum across operators in future networks.
\section{Conclusion}
\label{coclusion}
ONs, which make up the foundation of communication infrastructure, have emphasized high reliability and long-term operation. 
However, the rise of AI is reshaping their requirements and promoting the adoption of ONDT technologies.
This work proposed a GN-model-based ONDT architecture that enables flexible QoT management for each E2E light path. 
The proposed approach supports a stepwise evolution—from ring networks to data center interconnects and multi-operator connections—toward next-generation ONs.

\section*{Acknowledgments}
This work is supported in part by a research grant from Taighde Éireann (Research Ireland) under 22/FFP-A/10598 (Twilights), 18/RI/5721 (Open Ireland), 13/RC/2077 P2 at CONNECT: The Research Ireland Centre for Future Networks, and from the National Institute of Information and Communications Technology (NICT), Japan, under JPJ012368G50201 and JPJ012368C09001.

\ifCLASSOPTIONcaptionsoff
  \newpage
\fi



%\bibliographystyle{IEEEtran}
\bibliographystyle{IEEEtran}
\bibliography{references}
%\bibliography{IEEEabrv,references}

\begin{IEEEbiographynophoto}{Hideki Nishizawa}
is a Senior Research Engineer at NTT Network Innovation Laboratories, Japan. He received B.S. and M.S. degrees from Chiba University and a Ph.D. from Hokkaido University.
\end{IEEEbiographynophoto}
\begin{IEEEbiographynophoto}{Toru Mano}
is a research engineer at NTT Network Innovation Laboratories. He received his B.E. and M.E. degrees from the University of Tokyo and a Ph.D. from Hokkaido University.
\end{IEEEbiographynophoto}
\begin{IEEEbiographynophoto}{Kazuya Anazawa}
is a researcher at NTT Network Innovation Laboratories, Japan. He received his B.E. and M.E. degrees from the University of Aizu.
\end{IEEEbiographynophoto}
\begin{IEEEbiographynophoto}{Tatsuya Matsumura}
  received the B.E. and M.E. degrees in engineering from the Waseda University in 2019 and 2021, respectively.
  %He is a Researcher with the NTT Inc. Network Innovation Laboratories, Japan.
  In 2021, he joined NTT Laboratories, Yokosuka Japan. %His research interests include optical network design and autonomous control of optical networks.
\end{IEEEbiographynophoto}
\begin{IEEEbiographynophoto}{Takeo Sasai}
  received the B.E., M.E. and Ph.D. degrees from the University of Tokyo, Japan, in 2015, 2017, and 2025 respectively. In 2017, he joined NTT Laboratories, Yokosuka, Japan.
  %, where he works on coherent DSP for fiber-optic communications, pioneering optical network tomography and fiber-longitudinal performance monitoring.
\end{IEEEbiographynophoto}
\begin{IEEEbiographynophoto}{Masatoshi Namiki}
  is a Senior Manager at NTT Network Innovation Center, Japan. He received B.E. and M.E. degrees in Electronics from the Tokyo Institute of Technology in 2009 and 2011, respectively.
\end{IEEEbiographynophoto}
\begin{IEEEbiographynophoto}{Dmitrii Briantcev}
is a research fellow at the CONNECT Centre, Trinity College Dublin. He received a B.Sc. from Saint Petersburg State University and M.Sc. and Ph.D. degrees from KAUST.
\end{IEEEbiographynophoto}
  %is a research fellow at the CONNECT Centre, Trinity College Dublin. He received the B.Sc. in Radiophysics from Saint Petersburg State University (2018) and the M.Sc. (2020) and Ph.D. (2023) in Electrical and Computer Engineering from KAUST, Saudi Arabia.
  %He works on optical communications and digital twins, integrating simulation, data-driven channel modeling, and testbed validation.
%
%\begin{IEEEbiography}
%[{\includegraphics[width=1in,height=1.25in,clip,keepaspectratio]{authors/ambrosone.jpg}}]
\begin{IEEEbiographynophoto}{Renato Ambrosone}\;
is a Ph.D. student in the PLANET group at Politecnico di Torino. He earned an M.Sc. in Computer Engineering from Politecnico di Torino.
%His research focuses on digital twins for optical networks and on the control and orchestration of disaggregated optical infrastructures.
%, with QoT-aware decision making.
%\end{IEEEbiography}
\end{IEEEbiographynophoto}

\begin{IEEEbiographynophoto}{Esther Le Rouzic}
is a researcher at Orange Labs, France. She received her Telecommunication degree from Télécom Bretagne, an M.Sc. from University College London, and a Ph.D. from Télécom Paris.
\end{IEEEbiographynophoto}

  %received the Telecommunication degree from Télécom Bretagne, Bretagne, France, the M.Sc. degree from the University College London, London, U.K., in 1996, and the Ph.D. degree in electronics and communications from Télécom Paris, France, in 1999. She joined Orange Labs, Lannion, France, in 2000.
  %, where she has been involved in optical networks.
%
\begin{IEEEbiographynophoto}{Stefan Melin}
is a lead optical network architect at Telia Company. He holds a Master’s degree in Electronic Engineering with specialization in telecommunications and digital signal processing.
\end{IEEEbiographynophoto}
  %is a lead optical network architect at Telia Company and has over 20 years of experience in transmission systems and optical networks. He holds a Master’s Degree in Electronic Engineering with speciality on Telecommunications/digital signal processing.
  %His technical areas of expertise are techno-economic studies, network optimisation, procedures, rules and strategies.
  %for how to develop the Telia Company automated open optical networks.
%
\begin{IEEEbiographynophoto}{Oscar González-de-Dios}
is the Head of SDN deployments for Transport Networks at Telefonica Global CTIO. He received the M.S. and Ph.D. degrees from the University of Valladolid.
\end{IEEEbiographynophoto}
\begin{IEEEbiographynophoto}{Juan Pedro Fernandez-Palacios}
is leading the Metro and Core Transport Networks team at Telefonica Global CTIO. He received the M.S. degree from Polytechnic University of Valencia.
\end{IEEEbiographynophoto}
\begin{IEEEbiographynophoto}{Xiaocheng Zhang}
is a network architect at NTT DOCOMO BUSINESS, Inc. He received a B.E. degree in engineering from Shanghai Jiao Tong University in 2014.
%He has been engaged in the development of software-defined networking (SDN) technology and network designing method for open optical transport networks.
\end{IEEEbiographynophoto}
\begin{IEEEbiographynophoto}{Keigo Akahoshi}
is a researcher in photonic network operation automation at KDDI Corporation. 
He received B.E. and M.E. degrees from Kyoto University, Japan, in 2021 and 2023, respectively.
%His research interests include network arcihtecture and network service provisioning.
\end{IEEEbiographynophoto}
%
%\begin{IEEEbiography}
%[{\includegraphics[width=1in,height=1.25in,clip,keepaspectratio]{authors/ggrammel.png}}]
\begin{IEEEbiographynophoto}{Gert Grammel}\;
is a Principal Engineer at Juniper Networks (since 2011). He leads the Physical Simulation Environment (PSE) Group within the Telecom Infra Project (TIP).
%, where he initiated the development of GNPy, an open-source tool for optical network performance planning.
%His current work focuses on evolving GNPy into a digital twin for optical networks using AI/ML.
%\end{IEEEbiography}
\end{IEEEbiographynophoto}

%\begin{IEEEbiography}
%[{\includegraphics[width=1in,height=1.25in,clip,keepaspectratio]{authors/gborraccini.jpg}}]
\begin{IEEEbiographynophoto}{Andrea D'Amico}\;
Andrea D’Amico is a Researcher in the ONS Department at NEC Laboratories America. He holds a Ph.D. in Telecommunications Engineering from Politecnico di Torino. 
%At NEC, Andrea conducts advanced research for the integration of physical-layer modeling and system-level intelligence to support the evolution of large-scale optical networks. His work focuses on abstracting complex signal behaviors into models that inform automation, optimization, and control across diverse deployment scenarios. By aligning physical accuracy with network adaptability, he contributes to the development of flexible and high-capacity infrastructures capable of meeting the demands of next-generation connectivity.
%He leads the Physical Simulation Environment (PSE) Group within the Telecom Infra Project (TIP) since 2024.
%and he is an active developer of the open-source library GNPy since 2019.
%\end{IEEEbiography}
\end{IEEEbiographynophoto}

%\begin{IEEEbiography}
%[{\includegraphics[width=1in,height=1.25in,clip,keepaspectratio]{authors/gborraccini.jpg}}]
\begin{IEEEbiographynophoto}{Giacomo Borraccini}\;
currently holds the position of postdoctoral scientist at NEC Laboratories America. He received his Ph.D. in 2023 from the Politecnico di Torino.
%His research focuses on optical networks at the physical layer, mainly on optical transmission at the system level in terms of modeling and control.
%Since 2020 he is an active developer of the open-source library GNPy within the PSE subgroup of the Telecom Infra Project consortium.
%\end{IEEEbiography}
\end{IEEEbiographynophoto}

\begin{IEEEbiographynophoto}{Marco Ruffini}
is full professor at Trinity College Dublin, leading the OpenIreland lab and the Optical and Wireless networks research group.
%His main research is in the area of Intelligent networks, where he carries out pioneering work on the convergence of fixed-mobile and access-metro networks, AI-based control and digital twins of optical networks, quantum networking and fibre sensing. He has been invited to share his vision through several keynote and talks at major international conferences across the world.
%He authored over 220 international publications, 10 patents, contributed to industry standards and secured research funding for over € 14 milion, and contributed his novel virtual Dynamic Bandwidth Allocation (vDBA) concept to the BroadBand Forum standardisation body.
\end{IEEEbiographynophoto}

\begin{IEEEbiographynophoto}{Daniel Kilper}
is Professor of Future Communication Networks and Director of the CONNECT Centre at Trinity College Dublin, Ireland. He received M.S. and Ph.D. degrees in physics from the University of Michigan.
\end{IEEEbiographynophoto}

%\begin{IEEEbiographynophoto}{Daniel Kilper}
%is Professor of Future Communication Networks and SFI CONNECT Centre Director at Trinity College Dublin, Ireland.
%He holds an adjunct faculty appointment at Columbia University Data Science Institute and College of Optical Sciences, University of Arizona.
%He is CTO and co-founder of Palo Verde Networks, Inc., and is the Green Internet and Service Provisioning topical area editor for IEEE Transactions on Green Communications and Networking.
%He co-chairs the IEEE International Network Generations Roadmap (INGR) Optics Working Group.
%He received M.S. (1992) and Ph.D. (1996) degrees in physics from the University of Michigan.
%From 2000 to 2013, he was a member of the technical staff at Bell Labs.
% bio must be less than 150 words
%His work has been recognized with a NIST Communication Technology Lab Innovator Award and a Bell Labs President’s Gold Medal Award, and he served on the Bell Labs Presidents Advisory Council on Research.
%He holds 13 patents and has coauthored 6 book chapters and more than 170 peer-reviewed publications.
%His research is aimed at solving fundamental and real-world problems in communication networks, addressing interdisciplinary challenges for smart cities, sustainability, and digital equity.

%\begin{IEEEbiography}
%[{\includegraphics[width=1in,height=1.25in,clip,keepaspectratio]{authors/curri2.jpg}}]

\begin{IEEEbiographynophoto}{Vittorio Curri}
is Full Professor at Politecnico di Torino, Italy, where he leads the Open PLANET Lab. He received his Laurea and Ph.D. degrees from Politecnico di Torino. He is an IEEE, OPTICA, and AAIA Fellow.
\end{IEEEbiographynophoto}

%\begin{IEEEbiographynophoto}{Vittorio Curri}\;received the Laurea (cum laude) in Electrical Engineering (1995) and the Ph.D. in Optical Communications (1999) from Politecnico di Torino, Italy, where he is Full Professor in the Department of Electronics and Telecommunications and leads the Open PLANET Lab.
%He contributed to the development of the Gaussian Noise (GN) model for fiber propagation.
%His research focuses on fiber transmission modeling, optical link design, Raman amplification, and the role of the physical layer in open, disaggregated, and AI-driven optical networks.
%He is Scientific Chair of the GNPy open-source project within the Telecom Infra Project.
%He is IEEE, OPTICA, and AAIA Fellow.
%\end{IEEEbiography}

%\newpage


%\vfill



\end{document}